\documentclass{jpconf}
\usepackage{amssymb}
\usepackage{graphicx}
\usepackage{color}

\begin{document}

\title{Field-induced collective spin-exciton condensation in a quasi-2D d$_\mathrm{x^2-y^2}$-wave heavy electron superconductor}
\author{Vincent P. Michal}
\address{Commissariat \`a l'Energie Atomique,
INAC/SPSMS, 38054 Grenoble, France}
\ead{vincent.michal@cea.fr}

\begin{abstract}
The origin of the spin resonance observed in CeCoIn$_5$ with Inelastic Neutron Scattering is subject to debate. It has been shown recently that in this heavy electron compound at low temperature an instability to a ground state with coexisting d$_{x^2-y^2}$-wave superconductivity and Spin Density Wave (SDW) order in a magnetic field is a corollary of the consideration of a collective spin excitation mode in a quasi-2D d$_{x^2-y^2}$-wave Pauli-limited superconductor. This provides a natural scenario for the occurence of the puzzling high-field-low-temperature phase highlighted in CeCoIn$_5$.
We present perspectives on this ground state transition and propose directions for future experiment.
\end{abstract}

\section{Background information and model considerations in the spin-fermion approach}
Modelling and understanding collective phenomena in correlated electron systems in general and new superconductors in particular is an interesting and challenging issue. 
The two-side observation of spin resonances in high-temperature and heavy-fermion superconductors has raised the question of whether the phenomena in these very different systems have a common mechanism. 
Interpreting Inelastic Neutron Scattering data \cite{Stock} for CeCoIn$_5$ has recently given rise to interesting debate \cite{Chubukov2}. We have proposed \cite{Michal} that the ground state transition in magnetic field observed \cite{Kenzelmann} in superconducting CeCoIn$_5$ (superconducting critical temperature $T_c\simeq 2.3\mathrm{K}$) is a consequence of the existence of the superconductor collective spin excitation mode \cite{Stock} whose condensation\footnote{Here the term \emph{condensation} is used in the sense of a transition of a collective excitation mode to static ordering.} explains naturally the confinement of magnetic ordering within the superconductor.

Historically measurement in CeCoIn$_5$ phase diagram of a line of discontinuity in the specific heat indicated a second-order transition to a high-field-low-temperature phase ($10\mathrm{T}<H<11.5\mathrm{T}$, $T< 0.4\mathrm{K}$) which was initially thought to be a realisation of the Fulde-Ferrel-Larkin-Ovchinnikov (FFLO) state. Later observation of magnetic Bragg peaks \cite{Kenzelmann} at incommensurate wavector $\mathbf{q}=(1/2-\delta,1/2-\delta,1/2)$ (in units of the tetragonal crystal Brillouin zone dimensions) with $\delta\simeq0.05$ whose appearence coincided with this very transition line gave evidence that the ground state had a magnetic character with magnetic coherence length $\sim3000\AA$ \cite{Kenzelmann} extanding well beyond the vortex cores and ordered magnetic moments of $15\%$ the Bohr magneton \cite{Kenzelmann}. Nuclear Magnetic resonance (NMR) experiment \cite{Koutroulakis} supported characterise the magnetic phase.

\begin{figure}[t]
\centering
\includegraphics[width=16cm]{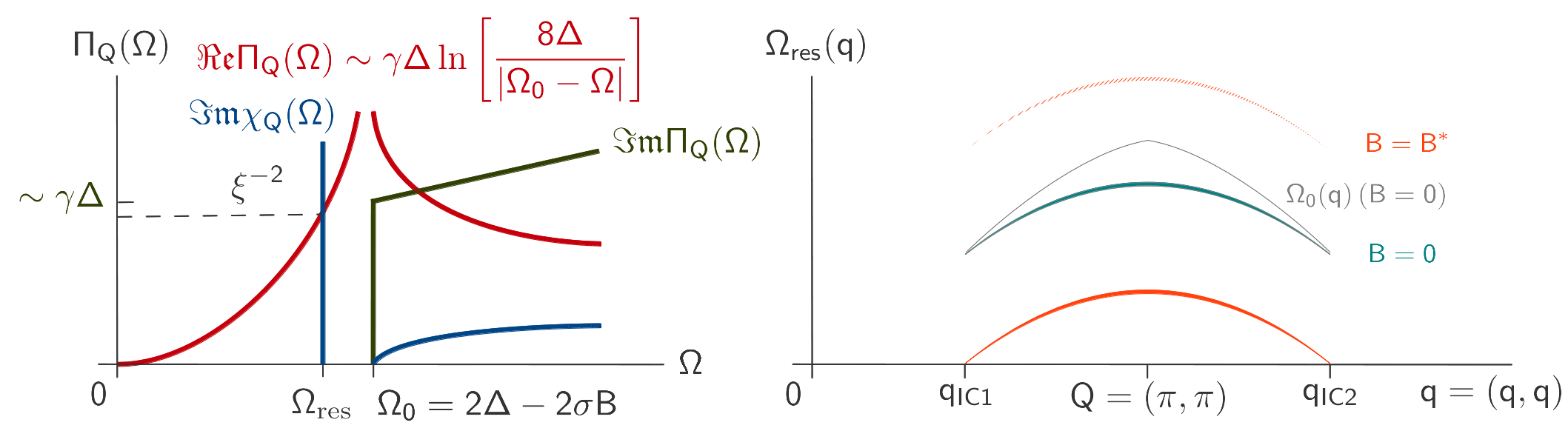}
\caption{Left: Behaviour of the uniaxial dynamic spin susceptibility (see text for discussion). The effect of the magnetic field is to split the threshold energy $\Omega_0$ into a two-step discontinuity. Right: Cartoon of the predicted resonant mode dispersion and threshold energy at zero field \cite{Chubukov1}, evolution with transverse field in the uniaxial case and criticality at incommensurate wave-vector.} 
\label{Image}
\end{figure}

When not superconducting CeCoIn$_5$ is a semi-metal with non-Fermi liquid resistivity exponent and a Kondo temperature $T_\mathrm{K}\sim7\textrm{K}$ while the Kondo lattice coherence temperature $T_{\mathrm{coh}}\sim50\textrm{K}$ \cite{Yang} yielding low temperature hybridized bands with large effective masses and a large specific heat discontinuity at the transition to the superconductor. In magnetic field the metal to superconductor transition is observed to be first order at temperature $T\lesssim 1\mathrm{K}$, a consequence of a large Maki parameter $\alpha_{M0}\sim H_\mathrm{c20}^\mathrm{orb}/H_\mathrm{c20}^\mathrm{p}$ ($H_\mathrm{c20}^\mathrm{orb}$ [$H_\mathrm{c20}^\mathrm{p}$] is the orbital [paramagnetic] upper critical field at zero temperature) which itself follows from a large electron effective mass. We consider a constant interaction magnetic instability from the superconducting, non-magnetic side. The model starts with a quasi-2D ($\alpha$-band in CeCoIn$_5$ \cite{q2D}) itinerant heavy electron band close to a commensurate SDW ground-state instability \cite{Pagliuso} with antiferromagnetic hot-spots (points in wave-vector space that belong to the Fermi line and are connected by the diagonal antiferromagnetic ordering wave-vector) which below $T_c$ becomes a superconductor with d$_{x^2-y^2}$-wave symmetry gap \cite{Vorontsov} and include a magnetic field $B$. In the spin-fermion approach \cite{Chubukov2, Chubukov1} spin density excitations and fermion excitations couple through a constant $g$ and yield the dynamic spin susceptibility
\begin{equation}
 \chi^{-1}(\mathbf{q},\Omega)=\chi_0^{-1}\Big[\xi^{-2}+|\mathbf{q}-\mathbf{Q}|^2-\Pi^{xx}(\mathbf{q},\Omega)\Big],
\label{Model}
\end{equation}
where $\xi$ is the magnetic correlation length.
To account for the magnetic anisotropy of the physical system, spin-density excitations are considered along a single axis (c-axis of the tetragonal crystal) whose direction is set prependicular to the magnetic field. The dimensionless spin excitation self-energy due to fermions (sometimes termed spin polarization function) $\Pi(\mathbf{q},\Omega)$ thus writes\footnote{Eq. \ref{Pi} includes magnetic field effect in the Pauli limit. The influence of the electron orbital motions on the spin susceptibility can be evaluated \cite{Eschrig} in the linearized Doppler shift approximation and gives correction of the order $B/(H_\mathrm{c20}^\mathrm{p}\alpha_{M0})$ which is small in a Pauli-limited superconductor as expected.}
\begin{eqnarray}
\nonumber\Pi^{xx}(\mathbf{q},\Omega)&=&-g^2\chi_0 T\sum_{\mathbf{k},m,\sigma}\Big[G_{\sigma}(\mathbf{k},i\omega_m)G_{-\sigma}(\mathbf{k}+\mathbf{q},i\omega_m+\Omega+i0^+)\\
&+&F_{\sigma}(\mathbf{k},i\omega_m)F_{-\sigma}^{+}(\mathbf{k}+\mathbf{q},i\omega_m+\Omega+i0^+)\Big],
\label{Pi}
\end{eqnarray}
where $G$, $F$ and $F^+$ are Green's functions in a superconductor with Zeeman field \cite{Michal} characterized by the spectrum of fermion excitations $E^{s}_{k\sigma}=s[\epsilon_k^2+\Delta_k^2]^{1/2}+\sigma B$ ($s,\sigma=\pm1$ are the particle-hole and spin indices respectively, $\epsilon_k$ is the zero-field spectrum in the metal \cite{Michal}, and the $\mathrm{d_{x^2-y^2}}$-wave gap in general reads $\Delta_k=(\Delta_0/2)\sum_na_n[\cos(nk_a)-\cos(nk_b)]$). The properties of the 2D function $\Pi$ at one loop level is known for zero field \cite{Chubukov2} and is considered here in Zeeman magnetic field (c. f. Fig. \ref{Image}). Here we consider that the dynamics of the mode is determined by the superconducting gap such that Eq. \ref{Model} yields the condition $\xi^{-2}+|\mathbf{q}-\mathbf{Q}|^2-\Re\mathfrak{e}\Pi(\mathbf{q},\Omega)=0$ ($\Omega>0$) defining the d$_{x^2-y^2}$-wave superconductor collective spin-excitation mode referred to as spin-exciton. The spectral function (the imaginary part of the dynamic spin susceptibility) of the mode looks like a Dirac delta in absence of Landau damping by particle-hole pair creation, which translates into the condition $\Im\mathfrak{m}\Pi(\mathbf{q},\Omega)/(\gamma\Delta)\ll1$ with $\gamma=Ng^2\chi_0/(2\pi|\mathbf{v}_\mathrm{khs}\times \mathbf{v}_\mathrm{khs+q}|)$, $N$ the number of hot-spots and $\mathbf{v}_\mathrm{khs}$ the Fermi velocity at hot-spot.

\section{From collective spin-excitation mode in a quasi-2D $\mathrm{\mathbf{d_{x^2-y^2}}}$-wave superconductor to ground state instability in a transverse magnetic field}
The fluctuation-dissipation theorem relates the dynamic structure factor measured in Inelastic Neutron Scattering experiment \cite{Stock,MagneticField} with the spectral function of the model (Eq. \ref{Model})
$\mathcal{S}(\mathbf{q},\Omega)=2\Im\mathfrak{m}\chi(\mathbf{q},\Omega+i0^+)/(1-e^{-\Omega/T})$,
the latter being the analytic continuation of the imaginary time spin correlation function $\chi(\mathbf{q},i\Omega_n)=\int_0^{1/T}d\tau\int  d^2\mathbf{r}\,e^{-i\mathbf{q}\cdot\mathbf{r}+i\Omega_n \tau}\langle S^x(\mathbf{r},\tau)S^x(0,0)\rangle$. 

\begin{figure}[t]
\centering
\includegraphics[width=16cm]{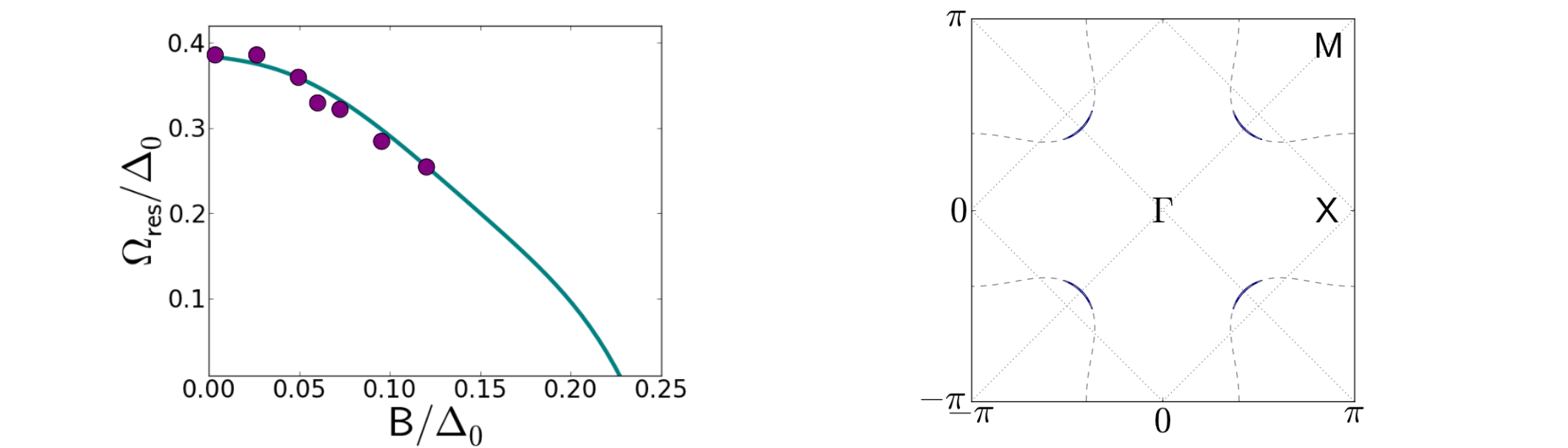}
\caption{Left: Collective mode energy variation with field at incommensurate wavevector $\mathbf{q}=(0.45,0.45)$, and comparison with experiment \cite{MagneticField}. Right: 2D Brillouin zone including in dotted grey the line where the superconducting gap changes sign and the antiferromagnetic reduced Brillouin zone, and in blue the Fermi pockets near the gap nodes at the condensation field $B^\ast=0.23\Delta_0$ \cite{Michal}.}
\label{Image2}
\end{figure}

The collective spin-excitation mode we discuss is a property of a quasi-2D d$_{x^2-y^2}$-wave superconductor on the border of a commensurate SDW continuous ground state transition
(an s-wave gap yields a negative $\Pi$ with no divergence and the logarithmic divergence (Fig. \ref{Image}) associated with the particle-hole excitation threshold in the d-wave case is due to two-dimensionality \cite{Chubukov2}). This d-wave effect is the one brought about by coherence factors in the BCS formalism. 

The mode dispersion is schematically represented in the right-hand side of Fig. \ref{Image} at zero field and at the condensation field. The mode spectral weight is maximum at commensurate wave-vector because there the threshold energy is also maximum. Away from this point the downward dispersion \cite{Chubukov1} follows from the wave-vector dependence of the d-wave gap in the Brillouin zone. A transverse magnetic field splits the mode into a branch which goes down in energy and an upper part which becomes damped by the continuum of particle-hole excitations. The consideration of magnetic isotropy gives a splitting between three undamped modes \cite{Ismer}. Ongoing Inelastic Neutron Scattering experiment \cite{MagneticField} is providing information on the actual magnetic anisotropy of the system and is showing deviation from strict uniaxiality.

The left hand side of Fig. \ref{Image2} shows the evolution of the resonance energy with field at fixed incommensurate wave-vector $\mathbf{q}=(0.45,0.45)$ \cite{Michal} where we have taken account of the variation of the gap magnitude with temperature and magnetic field by considering the superconductor gap equation (we found the zero temperature first order critical field $H_{c20}^{p}=0.36\Delta_0$ corresponding to the band structure considered in \cite{Michal}). The right-hand side of Fig. \ref{Image2} shows Fermi pockets induced by magnetic field approaching the antiferromagnetic hot-spot locations. Before the pockets reach these points the ground state transition occurs together with reconstruction of the Fermi surface. This must be visible with Nuclear Magnetic Resonance.

\section{Discussion, perspectives and conclusion}

We now turn to some perspectives on this instability of the superconducting ground state. First there is a possibility of a double-q structure  which follows from the mode dispersion (Fig. \ref{Image}) and is consistent with the prediction of Y. Kato et al. \cite{Kato}, there seen as a consequence of the incommensurate ordering wave-vector connecting nested pockets. As was emphasized in \cite{Kato}, this degeneracy is however expected to be lifted by small coupling between the electron orbital motion and the magnetic field. We suspect magnetostriction effects also play a part. 
It has been also conjectured \cite{Hatakeyama} the existence of a phase where SDW lives in a Larkin-Ovchinnikov (LO) superconducting state (where the order parameter spatially varies along the magnetic field direction). Because the two orders couple, this should give rise to additional length scale for space dependence of the SDW order parameter with experimental hallmark as Bragg peaks measurable with elastic neutron scattering. Such a prediction signalling coexistence between LO and SDW is waiting for experimental verification (this is also true for inhomogeneous superconductivity with a different spin state such as Pair Density Wave state). 

We saw that the specificities of the heavy electron superconductor CeCoIn$_5$ brings all conditions (first order transition \cite{Kenzelmann} resulting from Pauli limiting, tendency towards two-dimensionality \cite{q2D}, pairing symmetry \cite{Vorontsov}, border of antiferromagnetism \cite{Pagliuso}) for realizing a ground state instability which is the natural result of the condensation under magnetic field of a preexisting spin collective excitation mode in the d$_{x^2-y^2}$-wave superconductor. As noted in \cite{Chubukov2}, collective phenomena in heavy electron compounds involve a physics that stems from a lattice of localized f-electrons coupling with conduction electrons. Understanding the nature of the heavy delocalized electrons and their relation with superconductivity remains a significant challenge.
 
\section{Acknowledgments} 

It is a pleasure to thank V. P. Mineev for support and also D. Agterberg, J. Flouquet, S. Gerber, P. B. Littlewood, S. Raymond, and I. Vekhter for many stimulating discussions.\\

\end{document}